\documentclass[twocolumn]{aastex6}

\def\um{$\mu$m}
\shorttitle{Hard X-ray Emission from WDs}
\shortauthors{Chu et al.}
\begin{document}
\title{Hard X-ray Emission Associated with White Dwarfs. IV. Signs of Accretion from 
Sub-stellar Companions} 
%
%
\author{You-Hua Chu}
\affil{Institute of Astronomy and Astrophysics, Academia Sinica 
(ASIAA), No.1, Sec. 4, Roosevelt Road,
Taipei 10617, Taiwan, ROC}

\author{Jes\'us A.\ Toal\'a}
\affil{Instituto de Radioastronom\'{\i}a y Astrof\'\i sica (IRyA), 
UNAM Campus Morelia, Apartado postal 3-72, 58090 Morelia, Michoacan, Mexico}

\author{Mart\'\i n A.\ Guerrero}
\author{Florian F.\ Bauer}
\affil{Instituto de Astrof\'{\i}sica de Andaluc\'{\i}a,
 IAA-CSIC, c/Camino Bajo de Hu\'etor 50, E-18008 Granada, Spain}
 
 \author{Jana Bilikova}
 \affil{Astronomy Department, University of Illinois, Urbana, Illinois 61801, USA}
 
 \author{Robert A. Gruendl}
 \affil{Astronomy Department, University of Illinois, Urbana, Illinois 61801, USA}
 

\begin{abstract}
KPD\,0005+5106, with an effective temperature of $\simeq$200,000 K, is 
one of the hottest white dwarfs (WDs).  \emph{ROSAT} unexpectedly 
detected ``hard'' ($\sim$1 keV) X-rays from this apparently single WD. 
We have obtained \emph{Chandra} observations that confirm the spatial 
coincidence of this hard X-ray source with KPD\,0005+5106. 
We have also obtained \emph{XMM-Newton} observations of
KPD\,0005+5106, as well as PG\,1159$-$035 and WD\,0121$-$756,
which are also apparently single and whose hard X-rays were
detected by \emph{ROSAT} at 3$\sigma$--4$\sigma$ levels.
The \emph{XMM-Newton} spectra of the three WDs show remarkably 
similar shapes that can be fitted by models including a 
blackbody component for the stellar photospheric emission, a 
thermal plasma emission component, and a power-law component. 
Their X-ray luminosities in the $0.6-3.0$ keV band range 
from $4\times10^{29}$ to $4\times10^{30}$ erg~s$^{-1}$.   
The \emph{XMM-Newton} EPIC-pn soft-band ($0.3-0.5$ keV) lightcurve of 
KPD\,0005+5106 is essentially constant, but the hard-band ($0.6-3.0$ keV) 
lightcurve shows periodic variations. 
An analysis of the generalized Lomb-Scargle periodograms for the 
\emph{XMM-Newton} and \emph{Chandra} hard-band lightcurves finds 
a convincing modulation (false alarm probability of 0.41\%) with a 
period of 4.7$\pm$0.3 hr.  
Assuming that this period corresponds to a binary orbital period, 
the Roche radii of three viable types of companion have been 
calculated: M9V star, T brown dwarf, and Jupiter-like planet. 
Only the planet has a size larger than its Roche radius, 
although the M9V star and T brown dwarf may be heated by 
the WD and inflate past the Roche radius. 
Thus, all three types of companion may be donors to fuel 
accretion-powered hard X-ray emission.
\end{abstract}
%
%
%
%
\section{Introduction}    \label{sec:intro}     

Three types of X-ray sources are known to be associated with white 
dwarfs (WDs): (1) photospheric X-ray emission from a WD 
itself, (2) accretion of material from a close binary companion, 
as in cataclysmic variables, and (3) coronal X-ray emission from
a late-type binary companion, such as dMe stars.  The latter two 
types of sources require the WDs to be in binary systems and their
observed X-ray spectra commonly peak near 1 keV or higher, consistent with
thermal emission from plasma with temperatures of a few $\times$10$^6$ K.
In contrast, the photospheric X-ray emission from a WD is soft
and detectable only at photon energies $<$0.5 keV.
Examples of these different types of sources can be found in 
ROSAT and XMM-Newton archival studies of X-ray 
emission from WDs by \citet[][Paper I]{Odwyer03},
\citet[][Paper II]{Chuetal04b}, and \citet[][Paper III]{Betal10}.

Theoretically, cool WDs with convective envelopes can generate
magnetic fields to power coronae and become sources of X-ray 
emission \citep{Serber90,Thomas95}.
Observational searches for coronal emission above a cool WD's 
photosphere have thus far failed to provide convincing detections.  
A stringent upper limit on the X-ray luminosity of a single, cool, 
magnetic WD has been placed by a 31.8 ks Chandra
observation of GD\,356 to be $L_{\rm X} < 6.0\times10^{25}$ 
ergs s$^{-1}$ \citep{Weiss07}.

The above conventional wisdom was challenged by the detection of
hard (photon energy $\sim$ 1 keV) X-ray emission from apparently
single WDs.  The most outstanding cases have been
WD\,2226$-$216 \citep[the central star of the Helix Nebula;][]{Guerrero2001} 
and KPD\,0005+5106 (WD\,0005+511), which show a soft photospheric 
component and a distinct hard component peaking near 1 keV (Paper I).   
In the case of WD\,2226$-$216, sensitive Hubble Space Telescope 
observations were used to rule out any companion earlier than M5 
\citep{Ciardullo1999}; however, a recent analysis of its TESS lightcurve
suggested a $<$0.16 $M_\odot$ binary companion with an orbital 
period of 2.77 days \citep{Aller2020}.
In the case of KPD\,0005+5106, the lack of IR excess and H$\alpha$ 
emission associated with coronal activity has been used to exclude 
the existence of a late-type companion with a corona \citep{Chuetal04a}.

In Paper I and Paper III, similar hard component peaking near 1 keV has 
also been detected toward WD\,0339$-$451 at a 6$\sigma$ level, and 
WD\,0121$-$756 (aka WD\,0122$-$753J, RX\,J0122$-$7521), 
PG\,1159$-$035 (WD\,1159$-$34), 
and WD\,1333+510 at 3$\sigma$--4$\sigma$ levels; moreover, hard
X-ray emission manifested in the soft X-ray emission extending to harder
0.5--1.0 keV photon energy range is observed in WD\,1234+481 and 
WD\,1254+223.  
None of these WDs have known binary companion or show IR excess 
indicative of a late-type stellar companion.  Interestingly, at least four of these
WDs have stellar effective temperatures greater than 100,000 K -- 
KPD\,0005+5106, WD\,0121$-$756, PG\,1159$-$035, and WD\,2226$-$210.

Spectral and temporal properties of the hard X-ray emission from these
intriguing WDs may shed light on the origins of their hard X-rays; however,
the existing ROSAT Position Sensitive Proportional Counter (PSPC) 
observations are of inadequate quality.  For example, only $\sim$10 photons
with energies greater than 0.5 keV were detected in the ROSAT PSPC 
observations of WD\,0121$-$756 and PG\,1159$-$035 (Paper I, Paper III).
Better X-ray observations are needed.  

To investigate physical properties of hard X-ray emission from apparently
single WDs, we have obtained Chandra X-ray Observatory 
ACIS-S observation of KPD\,0005+5106, but its very 
luminous soft photospheric emission caused pile-up effects; therefore, we
have obtained XMM-Newton observations of KPD\,0005+5106 for
spectral and temporal analyses.  We have also acquired XMM-Newton 
observations of WD\,0121$-$756 and PG\,1159$-$035 to confirm their hard 
X-ray emission and to carry out spectral analyses.  This paper reports our
analyses of these Chandra and XMM-Newton observations:
Section 2 describes the X-ray observations as well as complementary IR
observations, Sections 3 and 4 report the spectral and temporal analyses of
the X-ray data, Section 5 discusses the implications of the X-ray results
on the physical origin of the hard X-ray emission, and finally Section 6
summarizes the conclusion of our study.

\section{Observations}     

\subsection{Chandra X-ray Observation}

The Advanced CCD Imaging Spectrometer (ACIS) on board the 
Chandra X-ray Observatory was used to observe 
KPD\,0005+5106 on 2008 March 19  (Obs.\ ID 8942; PI: Y.-H. Chu).
KPD\,0005+5106 was positioned at the aim-point of the ACIS-S array
on the back-illuminated CCD S3 and observed in the FAINT mode for 
a total of 48.0 ks.
The data were processed and analyzed using the Chandra 
Interactive Analysis of Observations (CIAO) software package 
\citep[version 4.11;][]{Fruscione2006}.
The observations were not affected by any period of high background, 
and no time intervals had to be excised.  
After dead time correction, the final exposure time was 47.4 ks.  

To refine the astrometric accuracy of this Chandra 
observation, we identified seven X-ray sources with optical 
counterparts and used their optical positions to calibrate the
X-ray astrometry.
The final astrometric accuracy of the Chandra observation
is better than 1$''$.
A bright X-ray point source is detected at the position of 
KPD\,0005+5106 with a background-subtracted count rate of 
0.157$\pm$0.002 cnts~s$^{-1}$ in the 0.15-4.5 keV energy band.

\begin{figure}
\figurenum{1}
\plotone{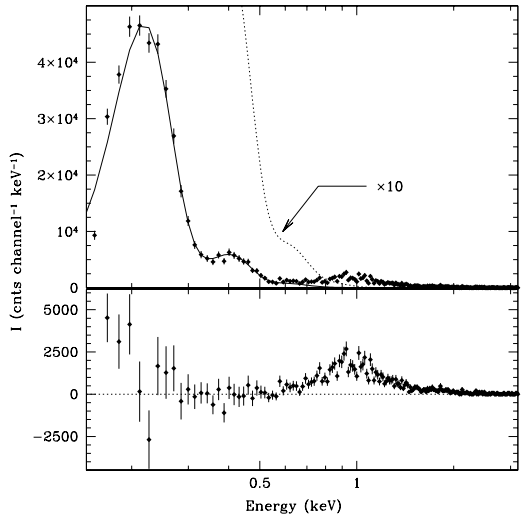}
\caption{Modeling the pileup in the Chandra ACIS-S 
spectrum of KPD\,0005+5106.  In the top panel, the data 
points with error bars show the background-subtracted raw 
spectrum, the solid curve shows a pileup model for the
spectrum, and the pileup model multiplied by a factor of 10
is plotted in a dotted curve to show the low-level effect
near 0.6 keV.  The bottom panel shows the pileup-removed
background-subtracted ACIS-S spectrum.
}
\label{fig:pileup}
\end{figure}

The background-subtracted ACIS-S spectrum of KPD\,0005+5106,
shown in Figure~\ref{fig:pileup}, exhibits  a bright peak near 
0.2 keV, a secondary peak near 0.4 keV and a much fainter 
tertiary peak near 0.6 keV.  
The bright primary peak corresponds to the soft photospheric 
emission from KPD\,0005+5106.
The very high count rate of the primary peak and the locations of 
the secondary and tertiary peaks suggest that the spectrum is 
affected by a pileup of photons with energies near the primary peak.

We have thus modeled the pileup effects following guidelines
in The Chandra ABC Guide to Pileup\footnote{
http://cxc.harvard.edu/ciao/download/doc/pileup$\_$abc.ps}
provided by the Chandra X-ray Center.
This model, plotted over the spectrum in Figure~\ref{fig:pileup},
indicates that the pileup contributions are still noticeable 
up to 0.75 keV.

The pileup-removed, background-subtracted spectrum of
KPD\,0005+5106, shown in the bottom panel of 
Figure~\ref{fig:pileup}, dips to nearly zero at 0.5 keV, peaks 
near 1 keV, and diminishes above $\sim$3 keV.  Its count rate is 
$\sim$0.024$\pm$0.001 cnts~s$^{-1}$ in the 0.6--4.5 keV band.
Because of the large uncertainties in Chandra ACIS-S 
calibration below 0.3 keV, the pileup correction is not ideal and 
the pileup-corrected spectrum of KPD\,0005+5106 would not be
suitable for spectral analysis.  We will nevertheless use the
Chandra ACIS-S data to extract lightcurves in the 
hard X-ray band ($\ge$0.75 keV, free of pileup effects) for 
comparisons with those extracted from the XMM-Newton
observations.

\begin{table*}
\centering
\caption{XMM-Newton Observations of Three Apparently Single WDs with Hard X-ray Emission}
\begin{tabular}{lccccccccccc}
    \hline
    \multicolumn{1}{c}{Object}&
    \multicolumn{1}{c}{Obs.\,ID.}&
    \multicolumn{1}{c}{Date}&
    \multicolumn{3}{c}{Total Exposure Time}&
    \multicolumn{3}{c}{Net Exposure Time}&
    \multicolumn{3}{c}{Source Count Rate}\\
    \cline{4-12}
    \multicolumn{1}{c}{}&
    \multicolumn{1}{c}{}&
    \multicolumn{1}{c}{}&
    \multicolumn{1}{c}{pn}&
    \multicolumn{1}{c}{M1}&
    \multicolumn{1}{c}{M2}&
    \multicolumn{1}{c}{pn}&
    \multicolumn{1}{c}{M1}&
    \multicolumn{1}{c}{M2}&
    \multicolumn{1}{c}{pn}&
    \multicolumn{1}{c}{M1}&
    \multicolumn{1}{c}{M2}\\
    \multicolumn{1}{c}{}&
    \multicolumn{1}{c}{}&
    \multicolumn{1}{c}{(yyyy-mm-dd)}&
    \multicolumn{1}{c}{(ks)}&
    \multicolumn{1}{c}{(ks)}&
    \multicolumn{1}{c}{(ks)}&
    \multicolumn{1}{c}{(ks)}&
    \multicolumn{1}{c}{(ks)}&
    \multicolumn{1}{c}{(ks)}&
    \multicolumn{3}{c}{(counts~ks$^{-1}$)}\\
    \hline
    KPD\,0005$+$5106 & 0693050401 & 2012-12-31 & 35.3 & 34.3 & 34.3 & 35.2 & 34.3 & 34.3 & 60.2 & 17.4 & 18.6\\
    WD\,0121$-$756   & 0693050501 & 2012-11-26 & 15.0 & 15.1 & 15.1 & 6.0  & 12.0 & 12.5 & 14.0 & 4.1  & 3.5 \\
    PG\,1159$-$035   & 0693050601 & 2012-12-12 & 10.0 & 10.1 & 10.1 & 5.0  & 9.5  & 9.8  & 7.5  & 2.3  & 1.2 \\
    PG\,1159$-$035   & 0723100301 & 2013-12-09 & \dots& 51.2 & 51.2 & \dots& 33.7 & 31.0 &\dots & 0.8  & 1.3 \\
    PG\,1159$-$035   & 0693050601 & 2019-01-04 & 24.2 & 26.0 & 26.0 & 18.3 & 25.2 & 24.2 & 5.8  & 1.7  & 1.2 \\
    \hline
\end{tabular}
\label{tab:observations}
\begin{list}{}{}
\item{Note: 
The three EPIC cameras are denoted as pn, M1, and M2.  The EPIC-pn observation 
0723100301 was performed in the small window mode and did not encompass PG\,1159$-$035.}
\end{list}
\end{table*}

\subsection{XMM-Newton X-ray Observations}

We obtained pointed XMM-Newton observations for  
KPD\,0005+5106, WD\,0121$-$756, and PG\,1159$-$035 
with the European Photon Imaging Cameras (EPIC) in 2012 
November and December (PI: Y.-H.\,Chu).
The Obs.\,IDs are 0693050401, 0693050501, and 0693050601, 
and the exposure times are 35.7~ks, 15.4~ks, and 10.4~ks, 
respectively. 
To avoid pileup, the Small Window mode and the 
Medium filter were used in the observations.  The Small Window mode 
reads out every  5.7 ms, and only count rates higher than 25 counts s$^{-1}$
cause pileup. 
In addition, we find in the XMM-Newton archive two 
observations that include PG\,1159$-$035 in the EPIC's 
field of view.   These two observations' target was the 
Seyfert-type galaxy Mrk\,1310, but PG\,1159$-$035 is 
located $\sim$8--9~arcmin from the aim point (depending on
the EPIC camera). Thus, these two observations, Obs.\ ID  
0723100301 (PI: N.\ Shartel) and 0831790501 (PI: B.\ Kelly), 
have been used to complement our observations of PG\,1159$-$035. 
Details of all the EPIC observations used in this paper are listed in
Table~\ref{tab:observations}, where the EPIC-pn, MOS1, and MOS2 
cameras are denoted as pn, M1, and M2, respectively.

The EPIC observations were processed with the XMM-Newton 
Science Analysis Software (SAS version 17.0) with the calibration 
files from 2019 June 13.  To avoid periods of high background, we
extracted lightcurves of the whole field in the 10--12~keV energy 
range using 100~s time bins, and excised time intervals with high
count rates in this high energy band.  The resultant useful times 
for each observation and camera are listed in Table~\ref{tab:observations}
as net exposure times.

Individual spectra were extracted from each of the pn and MOS observations
using the SAS task {\it evselect}.  To extract the spectra of a WD, a circular 
source aperture of 20$''$ and a surrounding or nearby background 
region devoid of sources  were used. 
Their corresponding calibration matrices, redistribution matrix file
(rmf) and auxiliary response file (arf),  were produced with the SAS tasks
{\it rmfgen} and {\it arfgen}.  Background-subtracted source count rates of
each camera obtained from different observations are listed in the
last three columns of Table~1.

EPIC spectra extracted from different cameras were combined to
produce a single EPIC spectrum for each WD. This has been done by
making use of the SAS task {\it epicspeccombine}. This task creates a
single background-subtracted spectrum for each source plus
corresponding calibration matrices\footnote{See the {\it
    epicspeccombine} thread in
  \url{https://www.cosmos.esa.int/web/xmm-newton/sas-thread-epic-merging}}. The
resultant combined EPIC spectra of KPD\,0005+5106, PG\,1159$-$035, and
WD\,0121$-$756, are presented in Figure~\ref{fig:spec}.

\subsection{Infrared Observations}

%

The Multiband Imaging Photometer for Spitzer \citep[MIPS;][]{Rieke04} 
on board the Spitzer Space Telescope was used to image KPD\,0005$+$5106 
in the 24 and 70~$\mu$m bands (Program ID 40953; PI: Y.-H. Chu). Images 
were obtained in photometry mode with the small offset scale and 10~s exposure 
time for three cycles at both 24 and 70~$\mu$m. The Basic Calibrated Data processed by the
Spitzer Science Center's pipeline software were used to construct 
mosaics in each band using utilities in the MOPEX software package.
Prior to building each mosaic, bad pixels and latent images of bright point
sources were flagged and removed, and the background brightness offsets 
between the individual frames were adjusted and removed.

KPD\,0005$+$5106 was not detected at either 24~$\mu$m or 70~$\mu$m.
We used the PHOT package in IRAF to obtain estimates for
the 3$\sigma$ detection limit in these two MIPS bands and find
flux density limits of $<$0.11 and $<$14.2 mJy at 24 and 70~$\mu$m, 
respectively.  These limits are too shallow to provide useful 
constraints on faint binary companions.

Spitzer InfraRed Array Camera \citep[IRAC;][]{Fazio04} 
observations of KPD\,0005+5106 were made in the 4.5 and 8.0 $\mu$m 
bands, and the flux densities were reported to be 297.5$\pm$9.9
and 96$\pm$14 $\mu$Jy, respectively \citep{Mullally07}.  These
values will be used in our analysis in Section 5.3.3.

Near-IR $JHK_s$ observations of KPD\,0005+5106 were also obtained
with the Near Infrared Camera Spectrometer (NICS) on the 3.58 m
Telescopio Nazionale Galileo (TNG) on 2008 October 12--13.
The detector was a HgCdTe Hawaii 1024$\times$1024 array with
a 4\farcm2$\times$4\farcm3 field of view.  The night was clear and
photometric.  KPD\,0005+5106 was observed at three epochs:
October 12 at UT = 19.4 h and October 13 at UT = 0.3 and 1.7 h.
At each epoch seven 50 s frames were obtained in each of the
$J$, $H$, and $K_s$ bands.  
No systematic variations are seen among the 21 measurements
in each band, and their averages are $J$ = 14.07$\pm$0.04,
$H$ = 14.15$\pm$0.03, and $K_s$ = 14.26$\pm$0.07.
These values can be compared with the 2MASS measurements
(2MASS J00081816+5123165) of $J$ = 13.95$\pm$0.03, $H$ = 
14.14$\pm$0.04, and $K_s$ = 14.19$\pm$0.06.
The $J$ magnitude shows a $\sim2\sigma$ difference, but the
$H$ and $K_s$ magnitudes are constant.
No large long-term variations are present.

\section{X-ray Spectral Analyses}      

\begin{figure*}
\figurenum{2}
\begin{center}
  \includegraphics[angle=0,width=\linewidth]{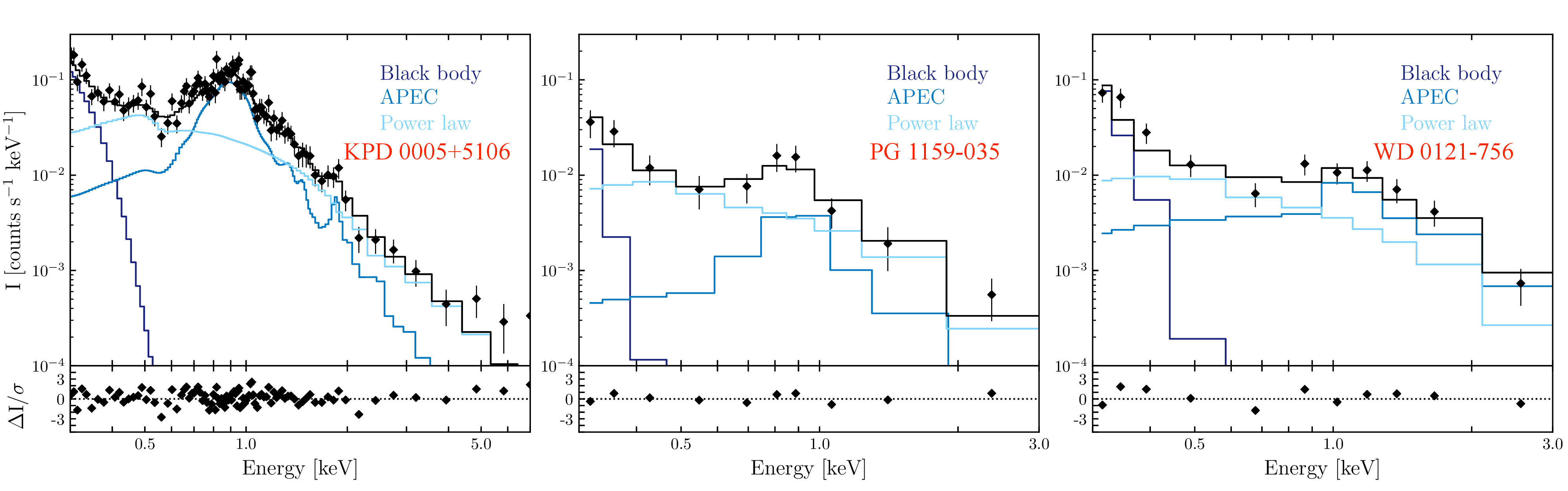}
\label{fig:spec}
\caption{XMM-Newton EPIC spectra of KPD 0005$+$5106,
  PG\,1159$-$035, and WD\,0121$-$756. The best-fit model to the data
  is plotted in solid black line. The contributions from the black
  body, {\it apec}, and power-law components are plotted in different shades 
  of blue lines.}
\end{center}
\end{figure*}

The XMM-Newton EPIC spectra, shown in Figure~\ref{fig:spec},
clearly detect the hard X-ray emission peak near 1 keV in all three WDs,
confirming the previous ROSAT PSPC's 3$\sigma$ detection at 
10 counts for PG\,1159$-$034, and WD\,0121$-$756.
The spectra of the three WDs are amazingly similar, showing a soft
component below 0.4 keV rising toward lower energies and a hard 
component peaking near $\sim$1 keV and extending to higher energies.
In the high-quality spectrum of KPD\,0005$+$5106, even some line features
are discernible in the 0.5 to 0.9~keV range.

\begin{table*}
\centering
\caption{Best-fit parameters to the EPIC spectra}
\begin{tabular}{lccccccccc}
    \hline
    \multicolumn{1}{c}{Object}&
    \multicolumn{1}{c}{$N_\mathrm{H}$}&
    \multicolumn{1}{c}{$T_\mathrm{eff}$}&
    \multicolumn{1}{c}{$kT$}&
    \multicolumn{1}{c}{$\Gamma$}&
    \multicolumn{1}{c}{$F_\mathrm{X,TOT}$}&
    \multicolumn{1}{c}{$L_\mathrm{X,TOT}$}&
    \multicolumn{1}{c}{$L_\mathrm{apec}/L_\mathrm{X,TOT}$}&
    \multicolumn{1}{c}{$L_\mathrm{pow}/L_\mathrm{X,TOT}$}\\
    \multicolumn{1}{c}{}&
    \multicolumn{1}{c}{($\times10^{20}$~cm$^{-2}$)}&
    \multicolumn{1}{c}{($\times10^{5}$~K)}&
    \multicolumn{1}{c}{(keV)}&
    \multicolumn{1}{c}{}&
    \multicolumn{1}{c}{(erg~s$^{-1}$~cm$^{-2}$)}&
    \multicolumn{1}{c}{(erg~s$^{-1}$~)}&
    \multicolumn{1}{c}{}&
    \multicolumn{1}{c}{}\\
    \hline
    KPD\,0005$+$5106 & 9$^{+5}_{-7}$ & {\bf 2.0}& 0.84$^{+0.07}_{-0.03}$ & 2.9$^{+0.4}_{-0.4}$  & $4.8\times10^{-13}$ & $8.7\times10^{30}$ & 0.18 & 0.27 \\
    PG\,1159$-$035   & {\bf 2.0}& {\bf 1.4}& 0.86$^{+0.45}_{-0.30}$       & {\bf 3.0}          & $1.3\times10^{-14}$ & $4.7\times10^{29}$ & 0.16 & 0.66 \\
    WD\,0121$-$756   & {\bf 4.0}& {\bf 1.8}& 2.00$^{+2.00}_{-0.65}$       & {\bf 3.0}& $9.2\times10^{-14}$ & $8.7\times10^{30}$ & 0.17 & 0.20 \\
    \hline
\end{tabular}
\label{tab:fits}
\begin{list}{}{}
\item{Note: Boldface text correspond to fixed values during the
  spectral fit. Fluxes and luminosities have been computed in the
  0.3--3.0~keV energy range for PG\,1159$-$035 and WD\,0121$-$756, but
  in the 0.3--7.0~keV for KPD\,0005$+$5106.}
\end{list}
\end{table*}

To gain insight into physical properties and origin of the X-ray emission 
from these apparently single WDs, we analyze their EPIC spectra with 
XSPEC \citep[version 12.10.1;][]{A96}.  We consider three types of 
emission -- blackbody model for stellar photospheric emission, thermal 
plasma emission model, and power-law model for nonthermal emission.  
All these emission models are absorbed using a {\it tabs} model as 
described in \citet{Wilms2000} and included in XSPEC. 
Initially, we model the spectra with two emission components, a 
blackbody component for the WD's photospheric emission and 
an optically thin plasma emission component or a nonthermal 
power-law component.  
None of the two-component models can successfully fit the EPIC 
spectra; they all result in reduced $\chi^{2}$ greater than 2.   
Thus, more complex models need to be considered.

As the EPIC spectrum of KPD\,0005$+$5106 has the highest 
quality and shows line features in hard X-rays 
(see Fig.\ \ref{fig:spec} - left panel), we start detailed spectral 
modeling of this spectrum by including (1) a blackbody 
component for the WD's photospheric emission with temperature 
fixed at the effective temperature of KDP\,0005$+$5106,
$T_\mathrm{eff}=2\times10^{5}$~K \citep{Wetal08,Wetal10}; 
(2) a thermal plasma emission
component for the line features; and (3) a thermal plasma 
or power-law component to improve the spectral fits.

For the three-component fits to KPD\,0005$+$5106's 
spectrum, we first consider another plasma emission model for the
third component, i.e., using the above fixed-temperature blackbody
emission model plus two optically thin {\it apec} plasma emission 
models.
The absorption column density as well as the plasma temperatures 
of the two {\it apec} models are left as free parameters. 
The best-fit model to the spectrum, with plasma temperatures of 
$kT_{1}$=0.20~keV and $kT_{2}$=0.62~keV and an unrealistically low 
absorption column density, has a reduced $\chi^{2}$ of 2.5 and hence 
is still not formally acceptable.

We then consider a power-law model for the third component, i.e., 
using a fixed-temperature blackbody component, an {\it apec} plasma 
emission component, and a power-law component.
The best-fit model to KPD\,0005+5106's spectrum has a plasma 
temperature of $kT$=0.84$^{+0.07}_{-0.03}$~keV, a power-law 
index of $\Gamma=2.9^{+0.4}_{-0.4}$, and an absorption column
density of $N_\mathrm{H}=(9^{+5}_{-7})\times10^{20}$ cm$^2$,
consistent with estimates from the Ly$\alpha$ absorption profile 
\citep{Wetal94}; furthermore, the reduced $\chi^{2}$ is improved to 
1.16.   We consider this best-fit model acceptable.  
The best-fit model's three individual emission components and 
their sum are plotted over the background-subtracted EPIC spectrum 
of KPD\,0005+5106 in the leftmost panel of Figure~\ref{fig:spec}.
It can be seen that the blackbody component of the photospheric 
emission dominates the spectrum at energies below 0.4 keV.  
The {\it apec} plasma emission component contributes to the broad 
peak around 1~keV and the emission line feature at $\sim$1.8~keV, 
which is likely a \ion{Si}{8} line. 
The power-law component dominates at high energies from 2~keV
up to 5--6~keV, as well as the intermediate energies at 0.4--0.6~keV.   
The intrinsic (unabsorbed) flux in the 0.3--7.0~keV energy range is 
$F_\mathrm{X}=4.8\times10^{-13}$~erg~s$^{-1}$~cm$^{-2}$, 
corrresponding to an X-ray luminosity of 
$L_\mathrm{X}=8.7\times10^{30}$~erg~s$^{-1}$ at the distance of 
387$\pm$8~pc adopted from the Gaia data release 2 \citep{Gaiadr22018}.
Detailed parameters of the best-fit model are given in Table~\ref{tab:fits}.

We note that multi-temperature plasma emission models also produce 
acceptable fits to the EPIC spectrum of KPD\,0005$+$5106. 
For example, we have fitted the EPIC spectrum with   a blackbody 
component plus three {\it apec} components. This results in a 
reasonably good fit ($\chi^{2}$=1.5) with plasma temperatures of
$kT_{1}=0.05^{+0.02}_{-0.007}$~keV, $kT_{2}=0.82^{+0.01}_{-0.05}$~keV, 
and $kT_{3}=2.6^{+1.0}_{-0.5}$~keV. 
The highest temperature component is needed for the X-ray emission
above $\sim$1.5 keV and the lowest temperature component is needed
for the X-ray emission around 0.4--0.6 keV.  The spectral fits can be 
further improved ($\chi^{2}$=1.1--1.3) if more plasma components are 
added.  Based on Occam's Razor, we do not pursue multi-temperature 
plasma model fits further; however, as we show in Section 4,  
multiple temperatures may be needed to explain the temporal variations 
of the spectral properties.  

Following our spectral analysis of KPD\,0005$+$5106, we find that
the spectra of PG\,1159$-$035 and WD\,0121$-$756 similarly need
to be modeled by a blackbody component, 
an {\it apec} thermal plasma component, and a power-law component.
These two WDs are fainter and have shorter exposure times; thus 
their observations detected much fewer counts than KPD\,0005+5106
and their spectra do not allow fitting many free parameters simultaneously.
We have converted these two WDs' extinction measurements into 
H column densities and adopt them as fixed absorption column densities
$N_\mathrm{H}$, which are $2\times10^{20}$~cm$^{-2}$ and 
$4\times10^{20}$~cm$^{-2}$  for PG\,1159$-$035 \citep{Dreizler1998} and
WD\,0121$-$756 \citep[][]{vanT1996}, respectively. 
The blackbody components' temperatures are fixed at the effective
temperatures of the WDs, $1.4\times10^{5}$~K for PG\,1159$-$035
\citep{Dreizler1998} and $1.8\times10^{5}$~K for WD\,0121$-$756
\citep{Werner1996}.
We also adopt the best-fit $\Gamma$ of KPD\,0005$+$5106 and fix the
$\Gamma$ of PG\,1159$-$035 and WD\,0121$-$756 at 3.

The resultant best-fit models for PG\,1159$-$035 ($\chi^{2}=1.20$) and
WD\,0121$-$756 ($\chi^{2}=1.9$) are listed in Table~\ref{tab:fits} and 
plotted over their background-subtracted EPIC spectra in Figure~\ref{fig:spec}.
It can be seen that the three components' contributions to PG\,1159$-$035 in
different energy ranges are very similar to those to KPD\,0005+5106.  
WD\,0121$-$756, on the other hand, appears to have the power-law component
dominating in the 0.4--0.7~keV energy range, while the plasma emission 
component dominates in the 1--3~keV energy range.  
Note that this dominance of  plasma emission in hard X-rays may be caused
by a higher plasma temperature in the best-fit model of WD\,0121$-$756, $kT$\,=\,2 keV,
instead of the $kT$\,$\sim$\,0.85 keV for KPD\,0005+5106 and PG\,1159$-$035.
Better spectral qualities are needed to confirm this result.  It is nevertheless interesting
to note that the power-law component  makes the most significant contribution to
the spectrum in the 0.4--0.6~keV energy range.

\begin{figure}
\figurenum{3}
\begin{center}
  \includegraphics[angle=0,width=\linewidth]{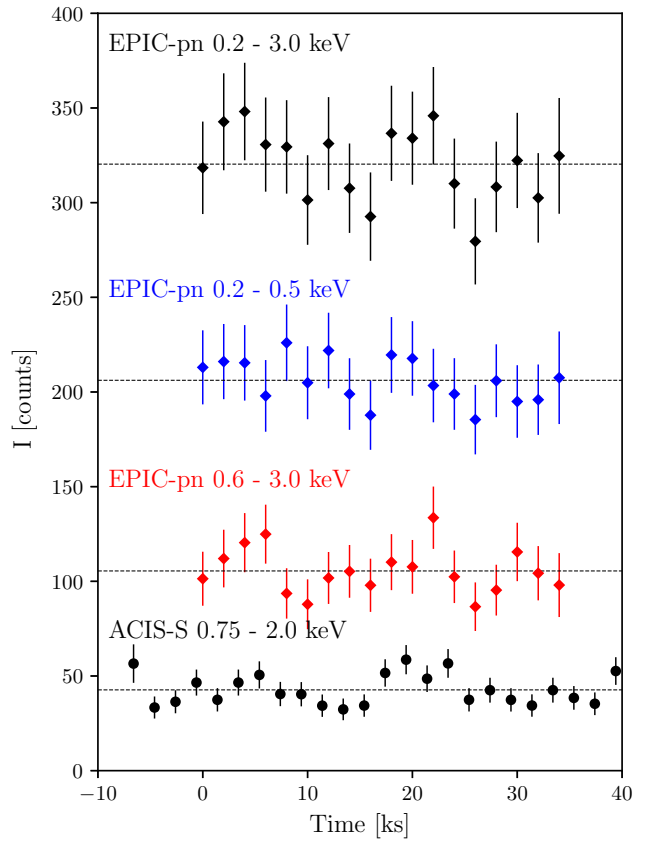}
\label{fig:lc}
\caption{Background-subtracted EPIC and ACIS-S lightcurves of
  KPD\,0005$+$5106. Each bin corresponds to 2~ks. The dashed thin
  lines represent the average values.}
\end{center}
\end{figure}

\section{Temporal Variations of Hard X-rays} 

Among the three WDs observed with XMM-Newton, 
only KPD\,0005$+$5106 had high enough a count rate and long enough 
an exposure time  (see Table~\ref{tab:observations}) to warrant temporal 
variation analyses.
We first extracted background-subtracted lightcurves from the EPIC pn
observation in three different energy ranges: the entire 0.2--3.0~keV 
energy band that covers the bulk of X-ray emission, the soft X-ray band
of 0.2--0.5~keV, and a hard band covering the 0.6--3.0~keV energy
range.  We have also extracted a lightcurve from the Chandra 
ACIS-S observation.  As stated in Section 2.1, the pileup effect is negligible
compared with the WD's hard X-ray emission at photon energies greater than
$\sim$0.7~keV.  Thus, we extracted a lightcurve in the 0.75--2.0~keV
energy range from the Chandra ACIS-S observations.  All four 
lightcurves are presented in Figure~\ref{fig:lc}.

The EPIC-pn lightcurves show a weak hint of variations in the broad band 
(0.2--3.0~keV) and essentially no variations in the soft band (0.2--0.5 keV);
however, variations are visible in the hard band (0.6--3.0~keV) and the 
two peaks are separated by $\sim$18 ks (= 5 hr).  
The ACIS-S lightcurve in the hard band (0.75--2.0~keV) shows similar
variations.  If the ACIS-S lightcurve is shifted to align its clearest peak
with the EPIC-pn hard-band lightcurve peak at 20 ks, it can be seen
that ACIS-S lightcurve also shows a lower peak at $\sim$4 ks mark 
where the EPIC-pn hard-band lightcurve peaks.  The similarity between
these two hard-band lightcurves lends support to the apparent 18 ks
period in the lightcurve.
To assess the significance of this periodic variation, we performed
Kolmogorov-Smirnov tests to the data and found the variation more
likely to be real than statistical fluctuations.  Thus, we performed more 
rigorous analysis using both ACIS-S and EPIC-pn data as described below.

\begin{figure*}
\figurenum{4}
\begin{center}
  \includegraphics[angle=0,width=\linewidth]{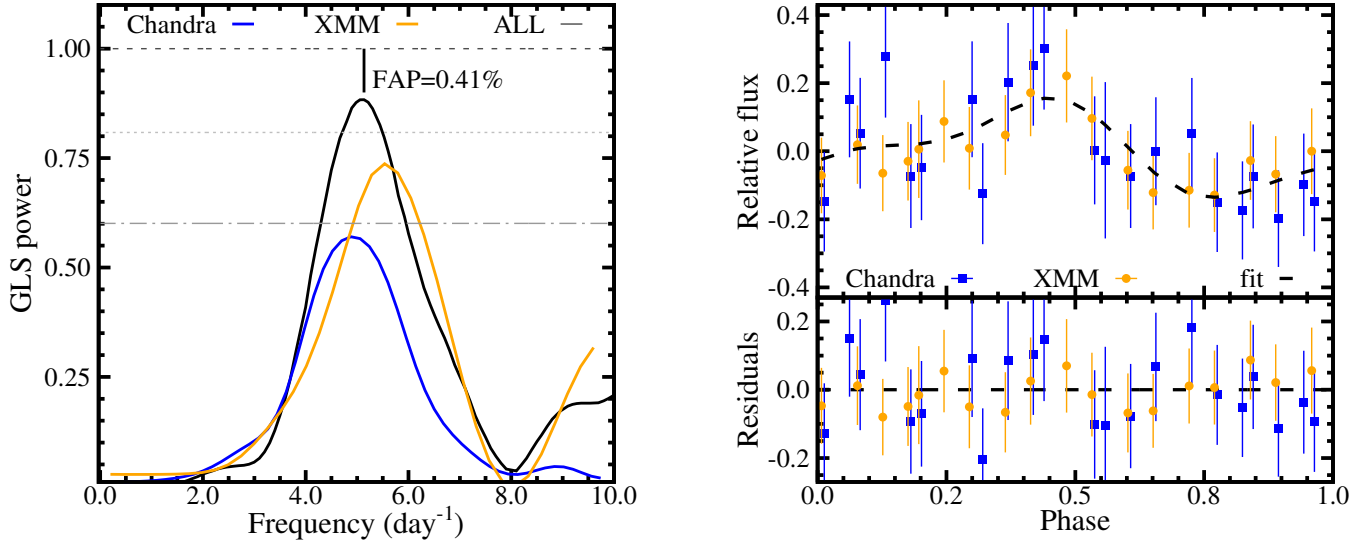}
\label{fig:periodogram}
\caption{Left panel: Generalized Lomb-Scargle (GLS) periodograms of Chandra
ACIS-S data (blue), XMM-Newton EPIC-pn data (orange),
and the combined data set (black). False alarm probabilities (FAP) of 10\%, 1\%,
and 0.1\% are indicated by dash-dotted, dotted, and dashed grey lines, 
respectively. Right panel: Phase-folded X-ray lightcurves of
Chandra ACIS-S data (blue squares), XMM-Newton EPIC-pn data 
(orange dots) and a spline model fit (black dashed line).}
\end{center}
\end{figure*}

We first normalized the Chandra ACIS-S and XMM-Newton
EPIC-pn hard-band lightcuves by their respective median flux values, 
then computed the generalized Lomb-Scargle periodograms 
\citep{Zechmeister2009} to find periodicities in these unequally spaced data 
and the false alarm probabilities (FAPs).  
We carried out computations for these two data sets both individually and combined.
The results are shown in the left panel of Figure~\ref{fig:periodogram},
with the generalized Lomb-Scargle periodograms of the Chandra data in blue,
XMM-Newton data in orange, and the combined data in black.
Both Chandra and XMM-Newton data independently show 
periodic variations close to 5 hr (frequency $\sim$5 day$^{-1}$), although 
each data set covers only 2-3 cycles of this modulation.
The combined data set covers 4.6 cycles, and its X-ray intensity modulation 
for a period of 4.7$\pm$0.3 hr should be more robust.  The error bar is
estimated at 1\% FAP.
Indeed, the FAP decreases from 10\% for the Chandra
data and 2\% for the XMM-Newton data, to 0.41\% for the combined
data set.  
The Chandra and XMM-Newton lightcurves folded to 4.7 hr,
displayed in the right panel of Figure~\ref{fig:periodogram}, 
show similar shapes within the error limits.

\begin{figure}
\figurenum{5}
\begin{center}
  \includegraphics[angle=0,width=\linewidth]{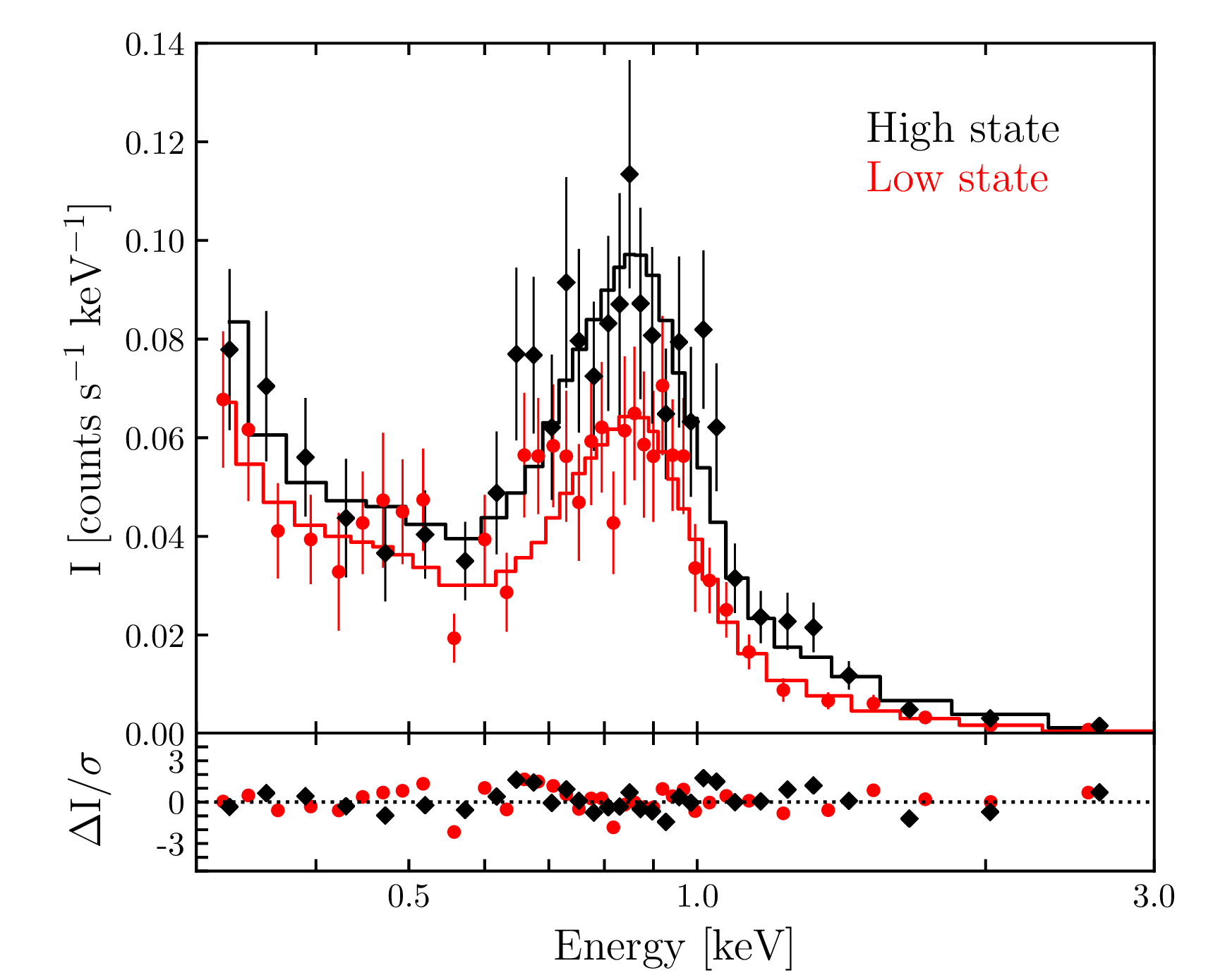}
\label{fig:spec_compare}
\caption{Background-subtracted EPIC pn spectra extracted from time
  intervals in the high and low states, plotted in black and red, respectively.}
\end{center}
\end{figure}

To investigate whether the X-ray spectral properties vary between
the high state and low state of KPD\,0005+5106, we have extracted 
an EPIC-pn spectrum from time intervals near the peaks in the 
hard-band lightcurve, and another EPIC-pn spectrum from time 
intervals near the valleys of the hard-band lightcurve.
The two spectra, shown in Figure~\ref{fig:spec_compare},
 are statistically similar at energies below 0.5~keV, 
but the high-state spectrum is brighter than the low-state spectrum, 
which is not surprising as the high-state and low-state spectra 
correspond to peaks and valleys of the lightcurve, respectively.

To quantify the differences between the high- and low-state EPIC-pn
spectra, we have modeled them with three components (blackbody +
thermal plasma + power law) as discussed in length in Section ~3.
Although the resultant best-fit models for the high- and low-state spectra
are suggestive of small variations in the plasma temperature and the 
power-law photon index $\Gamma$, they are also consistent within 
their error bars.  It is nevertheless clear that the best-fit models indicate 
the high-state is 1.5 times as bright as the low-state regime.

We note that in addition to the obvious luminosity differences, there
are subtle differences in the line features. For example, compared with the 
low-state spectrum, the high-state spectrum shows a stronger contribution 
from Ne lines at $\sim$1.0~keV and an excess at $\lesssim$1.4~keV 
probably due to the Mg\,{\sc xi} line emission at a 2--3$\sigma$ confidence level.
Moreover, the low-state spectrum exhibits
an emission peak near 0.5~keV that is also present in the EPIC spectrum
in Figure~\ref{fig:spec} but not seen in the high-state spectrum.
It would be very interesting to analyze detailed differences 
between the high- and low-state spectra, for example, plasma
temperatures and abundances, the number of thermal plasma 
components needed in spectral fits, and their relative 
contributions (see Section~3); however, the 
current data do not have adequate counts and resolution for such 
analyses.  Future deeper observations would be desirable.

\section{Discussion}

We first use the positional coincidence to affirm the physical
association between the hard and soft X-ray emission components
in the WDs, then compare the spectral properties among the three WDs,
and discuss the possible origins of the hard X-ray emission.

\subsection{Spatial Coincidence between the Hard and Soft X-ray Emission}

The superb angular resolution of Chandra enables us to 
determine accurately whether a position offset exists between 
the hard and soft X-ray emission from KPD\,0005+5106.  
We have extracted images of KPD\,0005+5106 in a soft band 
(0.2--0.5~keV) and a hard band (0.8--3.5~keV), and find that 
the soft and hard X-ray point sources are coincident with each other  
(in the same 0\farcs5 pixel) and with the optical position of the WD 
(within 1$''$).  These coincidences support that both the soft
and hard X-ray components are attributed to KPD\,0005+5106.

The angular resolution of XMM-Newton is not as good 
as Chandra, yet it can still be shown from the pointed 
observations of PG\,1159$-$035 and WD\,0121$-$756 that both
their hard and soft X-ray components and the optical position
of the WD are coincident with one another within $\sim$1$''$.
We therefore consider that the spatial coincidences affirm the physical
association between the hard and soft X-ray components and between
the X-ray emission and the WD.  

\subsection{Similarities among the Three WDs}

KPD\,0005+5106 is a DO WD, while PG\,1159$-$035 and 
WD\,0121$-$756 are PG\,1159 type WDs.  Both DO and
PG\,1159 spectral types imply a H-deficient, He-rich
atmosphere.  Besides the similarity in composition, all three
have stellar effective temperatures greater than 100,000 K:
200,000 K, 140,000 K, and 180,000 K for KPD\,0005+5106,
PG\,1159$-$035, and WD\,0121$-$756, respectively.

The X-ray spectral properties of the three WDs are also
similar.  As analyzed in Section 3, all three WDs have similar
X-ray spectral shapes.  They all possess a soft X-ray component 
corresponding to the stellar photospheric emission and hard
X-rays peaking near 1 keV.  Furthermore, the hard X-ray 
emission cannot be fitted well by two thermal plasma 
emission components; instead, it requires models 
consisting of a thermal plasma emission component and a 
power-law component (or multi-temperature plasma components).

These similarities in physical properties of the WDs and in
X-ray spectral properties suggest that all three WDs may 
share a common origin for their hard X-ray emission.

\subsection{Origin of the Hard X-ray Emission}

We will consider three possible origins of the observed hard X-rays
from these apparently single WDs:
(1) photospheric emission, (2) hidden coronal companion, and 
(3) accretion from a hidden companion.

\subsubsection{Photospheric Emission} 

KPD\,0005+5106 is the most well-studied and its X-ray spectrum
has the highest quality among the three WDs reported in this paper.  
We will thus first examine previous studies of KPD\,0005+5106 in detail. 

KPD\,0005+5106 is such a bright soft X-ray source that it was 
detected in the ROSAT All Sky Survey.
\citet{Fleming93} showed that the spectrum of KPD\,0005+5106 in 
the 0.1--0.4 keV range could be fitted by a model of thermal plasma
at a temperature of 2.6$\times$10$^5$ K, and suggested that its 
X-ray emission originated from a corona cooler than those of typical
main-sequence stars.
A corona for KPD\,0005+5106 would be surprising because the star is 
fully ionized and should not have a convective envelope to power a 
corona.
Indeed, based on a Chandra Low Energy Transmission Grating 
Spectrograph (LETGS) observation, \citet{DW05} found that the X-ray
spectrum of KPD\,0005+5106 at 20--100 \AA\ (0.12 -- 0.62 keV) could 
be better modeled as photospheric emission with effective temperature 
of 120,000 K and log $g$ = 7.
They ruled out coronal emission as the X-ray source
because the extrapolated optical continuum intensity would exceed 
the observed photospheric continuum, and because of a lack of 
H-like and He-like C lines that should otherwise have been observed
in X-rays.  
Their 120,000 K photosphere model can explain the
soft X-ray spectrum, but cannot reproduce the hard X-ray
emission from KPD\,0005+5106.

The discovery of photospheric \ion{Ca}{10} emission lines 
in the Far Ultraviolet Spectroscopic Explorer (FUSE)
spectrum, in conjunction with the \ion{Ne}{8} lines in UV and 
optical spectra, provided evidence that KPD\,0005+5106
has an effective temperature of $\sim$200,000 K, much hotter
than previous estimates \citep{Wetal07,Wetal08}.
In light of this higher temperature, the stellar spectrum of 
KPD\,0005+5106 has been re-modeled by \citet{Wetal10}.
The new model not only has a higher temperature but also higher 
metallicities than those in the model of \citet{DW05}.
Despite its higher temperature, this new model cannot reproduce
the observed hard X-ray emission from KPD\,0005+5106. 

We conclude that the hard X-ray emission from KPD\,0005+5106 
does not have a photospheric origin.  This conclusion may apply to
PG\,1159$-$035 and WD\,0121$-$756 as well.

\subsubsection{Hidden Companions with Active Coronae}

\begin{deluxetable*}{crrrrrr}
\label{tab:mk_lx}
\tablewidth{0pt}
\tablecaption{Infrared and X-ray Properties of Putative Late-type Dwarf Companions}
\tablehead{
\multicolumn{1}{c}{Spectral Type} &
\multicolumn{2}{c}{\underline{~~~~~~KPD\,0005$+$5106~~~~~~}}  &
\multicolumn{2}{c}{\underline{~~~~~~PG\,1159$-$035~~~~~~}}  &
\multicolumn{2}{c}{\underline{~~~~~~WD\,0121$-$756~~~~~~}}  \\
\multicolumn{1}{c}{} &
\multicolumn{1}{c}{$m_k$} & \multicolumn{1}{c}{$\log (L_{\rm X}/L_{\rm bol})$} &
\multicolumn{1}{c}{$m_k$} & \multicolumn{1}{c}{$\log (L_{\rm X}/L_{\rm bol})$} &
\multicolumn{1}{c}{$m_k$} & \multicolumn{1}{c}{$\log (L_{\rm X}/L_{\rm bol})$}
}
\startdata
K0 & 11.90~~~~~ & $-$2.62~~~~~~~~~~ & 12.64~~~~~ & $-$3.62~~~~~~~~ & 13.69~~~~ & $-$2.70~~~~~~~~ \\
K5 & 12.46~~~~~ & $-$2.17~~~~~~~~~~ & 13.20~~~~~ & $-$3.17~~~~~~~~ & 14.25~~~~ & $-$2.26~~~~~~~~ \\
M0 & 13.11~~~~~ & $-$1.88~~~~~~~~~~ & 13.85~~~~~ & $-$2.88~~~~~~~~ & 14.90~~~~ & $-$1.97~~~~~~~~ \\
M5 & 14.09~~~~~ & $-$1.03~~~~~~~~~~ & 14.83~~~~~ & $-$2.03~~~~~~~~ & 15.88~~~~ & $-$1.12~~~~~~~~ \\
M8 & 15.26~~~~~ & $-$0.07~~~~~~~~~~ & 16.00~~~~~ & $-$1.07~~~~~~~~ & 17.05~~~~ & $-$0.16~~~~~~~~ \\
WD & 14.19$\pm$0.06 &    &  15.73$\pm$0.23  &   &  16.21$\pm$...~ &   \\
\enddata
\end{deluxetable*}

Previously, \citet{Chuetal04a} assumed the hard X-ray emission of
KPD\,0005+5106 originated completely from a coronal companion, and 
showed that such companion would outshine the WD itself in near-IR
passbands.  Here we use a similar but more systematic approach 
to examine the possible presence of hidden companions.

To assess the possibility that the hard X-ray emission from KPD\,0005$+$5106,
PG\,1159$-$035, and WD\,0121$-$756 originates from the coronal activity of
late-type dwarf companions, we first compare the brightness of the WD with
a potential companion in the $K$ band to see whether a companion can hide
underneath the bright WD emission. 
We have computed the expected $m_K$ magnitudes of K0, K5, M0, M5, and M8 
dwarf stars at the distances of KPD\,0005$+$5106 (390 pc), PG\,1159$-$035 
(550 pc), and WD\,0121$-$756 (890 pc), using the standard star magnitudes
from \citet{SK1982} and distances from Gaia data release 2 \citep{Gaiadr22018}.
The expected $m_K$ magnitudes of these stars, as well as the observed
$K$ magnitudes the WDs, are listed in  Table 3. 
Obviously, a bright companion cannot hide behind a fainter WD; thus only
the latest type of M dwarfs are candidates of hidden companions.

We next consider whether coronae of late-type companions are able to provide
the hard X-ray luminosities of these three WDs: 
3.9$\times$10$^{30}$, 3.9$\times$10$^{29}$, and
3.2$\times$10$^{30}$ erg~s$^{-1}$ for KPD\,0005+5106, PG\,1159$-$035,
and WD\,0121$-$756, respectively.
We computed $\log (L_{\rm X}/L_{\rm bol})$ for the putative late-type dwarf 
companions, and listed them in Table 3, too. 
The highest $\log (L_{\rm X}/L_{\rm bol})$ observed in K--M dwarf stars with
saturated coronal activity is about $-$3.5, and some flare stars can reach $-$2.8
 \citep{Fetal95,Gudel04}. 
The values of $\log (L_{\rm X}/L_{\rm bol})$ listed in Table 3 indicate that 
these late type dwarf stars would have $\log (L_{\rm X}/L_{\rm bol})$ exceeding
the range, up to about $-$3, that can possibly be provided by 
stellar coronae.

\begin{figure}
\figurenum{6}
\begin{center}
  \includegraphics[angle=0,width=\linewidth]{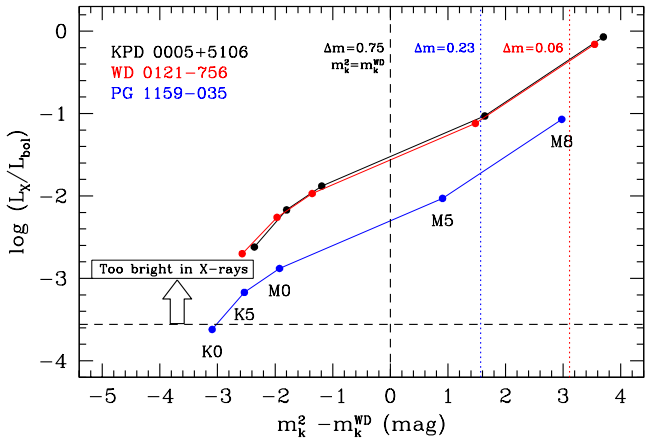}
\label{fig:Lx_Mk}
\caption{X-ray to bolometric luminosity ratio and infrared excess in the $K$ band of putative
late-type dwarf companions of KPD\,0005$+$5106 (black), PG\,1159$-$035 (blue), and
WD\,0121$-$756 (red). 
The different spectral types have been connected by dotted lines for each WD and
are labeled in the track of PG\,1159$-$035. 
Dots to the left of the vertical dashed line imply that
putative late-type dwarf companions would be actually
brighter than the observed $m_K$ magnitudes of
KPD\,0005$+$5106 (14.26),
PG\,1159$-$035 (15.733), and
WD\,0121$-$756 (16.213). 
Dots above the horizontal dashed line imply that putative late-type dwarf companions
would have actually X-ray to bolometric luminosity ratios in excess of the canonical
value for saturated activity in these stars.}
\end{center}
\end{figure}

Figure~\ref{fig:Lx_Mk} is another presentation comparing the brightness 
and $\log (L_{\rm X}/L_{\rm bol})$ of putative late-type dwarf companions
with those of the WD.  All K0--M8 dwarf stars are either too bright in the
$K$ band and/or cannot provide the observed hard X-ray luminosities.
Therefore, we can rule out coronal activities of a late-type companion
for the origin of the hard X-ray emission detected in KPD\,0005$+$5106,
PG\,1159$-$035, and WD\,0121$-$756.

\subsubsection{Accretion from a Hidden Companion}

Finally we consider accretion-powered hard X-ray emission from the WDs.
We will use KPD\,0005+5106 as an example for our analysis, as it has the 
most well determined physical parameters among the apparently single WDs 
with hard X-ray emission.  
We adopt a mass of 0.64 $M_\odot$ and a radius of 0.059 $R_\odot$ (or
6.5 $R_\oplus$) for KPD\,0005+5106 \citep{Wetal10}.

\begin{figure}
\figurenum{7}
\begin{center}
  \includegraphics[angle=0,width=\linewidth]{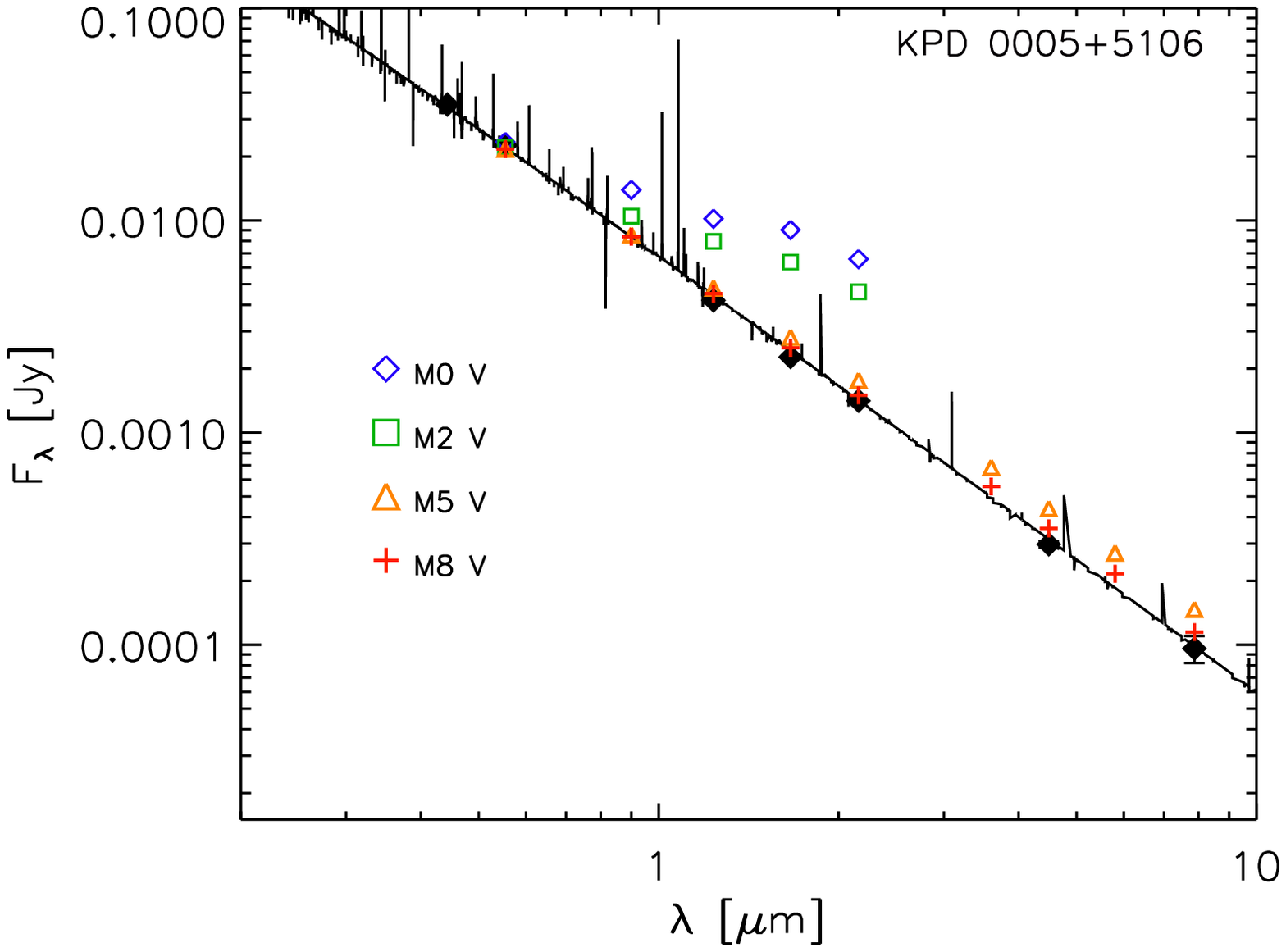}
\label{fig:KPD_sed}
\caption{Spectral energy distribution (SED) of KPD\,0005+5106.  
The solid symbols are data: $B$ and $V$ \citep{Detal85,MS99}, 2MASS $J$, $H$,
and $K_s$,  Spitzer IRAC 4.5 and 8 $\mu$m \citep{Mullally07}.
The error bars are all smaller than the symbols, except the 8\um\ measurement.
The open symbols are models assuming M-type companions.  The photometry
of standard M stars are from \citet{KM94} and \citet{Petal06}.
The solid curve is a model SED of KPD\,0005+5106 from \citet{Wetal10}
normalized to the $K_s$ band.}
\end{center}
\end{figure}

We use the Spitzer IRAC observations and the 2MASS, or TNG 
NICS, $JHK_s$ observations of KPD\,0005+5106 to place constraints on
possible companions. 
The $B$ and $V$ magnitudes of KPD\,0005+5106 have been measured
by \citet{Detal85}: $B$ = 13.02 and $V$ = 13.32.
These optical and IR photometric measurements are used to produce 
the spectral energy distribution (SED) of KPD\,0005+5106 shown 
in Figure~\ref{fig:KPD_sed}, where the model spectrum from \citet{Wetal10}
is also plotted. 
The extinction correction has been made by using the hydrogen column
density of $N_{\rm H}$ = $5\times10^{20}$ H-atoms cm$^{-2}$ toward 
KPD\,0005+5106 \citep{Wetal94} and the gas-to-dust ratio of 
$N_{\rm H}$/$E(B-V)$ = $5.8\times10^{21}$ H-atoms cm$^{-2}$ mag$^{-1}$
\citep{Betal78}.  

Comparing the observed SED with these model SEDs (Fig.~\ref{fig:KPD_sed},
see figure caption for origin of data and references), we can easily
rule out the existence of a companion of spectral type M5 or earlier.
The observed 4.5 and 8.0 \um\ fluxes are $\sim$20\% lower than those
expected with an M8\,V companion.  
Suffice it to say, any hidden companion of KPD\,0005+5106 has to be 
fainter and less massive than an M8\,V star.

We will consider three types of hidden companions, an M9\,V star with a mass 
of 0.075 $M_\odot$ and a radius of 0.08 $R_\odot$ \citep{KT2009}, 
a T type brown dwarf with a 
mass of 0.035 $M_\odot$ and a radius of $\sim$0.1 $R_\odot$, and a 
Jupiter-like planet with a mass of 0.001 $M_\odot$ and a radius of 
$\sim$0.1 $R_\odot$.
We assume that the period of 4.7 hr in the lightcurve of
KPD\,0005+5106's hard X-ray emission is the orbital period of this binary 
system. 
Using the Kepler's third law, we can determine the separation between the
WD and the companion, 1.27 $R_\odot$ for a M9\,V star,  1.24 $R_\odot$ 
for a T brown dwarf, and 1.22 $R_\odot$ for a Jupiter-like planet.
The effective Roche radius $r_L$ of the companion can be approximated by: 
\begin{equation}
r_L/a = \frac{0.49~q^{2/3}}{0.6~q^{2/3} + {\rm ln}~(1 + q^{1/3})}~,
\end{equation}
where $a$ is the separation between the WD and the companion and
$q$ is the companion to WD mass ratio \citep{Eggleton1983}.
We find Roche radii of 0.28 $R_\odot$, 0.21 $R_\odot$, and 
0.067 $R_\odot$ for a M9\,V star, a T brown dwarf, and a Jupiter-like planet, 
respectively.  
Comparing these Roche radii with their respective radii, it is clear that a 
Jupiter-like planet is larger than the Roche radius and can channel mass 
to the WD
easily.  The atmospheres of M9\,V star and T brown dwarf, being 
ionized and heated 
by the 200,000 K KPD\,0005+5106 at a distance of $\sim$1.3 $R_\odot$, 
may be inflated and the outer edge of their atmospheres may exceed
the Roche radius and be accreted by the the WD. 
We conclude that all three types of hidden companions may be the donor
providing material for accretion-powered hard X-ray emission, although 
a Jupiter-like planet can do it most easily.

Previously, $\sim$10 WDs accreting from brown dwarfs have been reported 
\citep{Longstaff2019}.  Only two of them show X-ray emission, 
EF Eridani \citep{Schwope2007} and 
SDSSJ121209.31+013627.7 \citep{Stelzer2017};
however, both are cool WDs with effective temperatures lower than 10,000 K,
and both are polars whose strong magnetic fields channel material from
the companion to the WDs' magnetic poles to be accreted.
The three WDs with hard X-ray emission reported in this paper 
have much higher effective temperatures than these polars.
As noted above, these hot WDs can accrete material from either a Jupiter-like
planet or an irradiated brown dwarf, and hence are very different from the
polars.

If we assume that the hard X-ray emission is powered by accretion from a 
companion, then the accretion rate would be $L_\mathrm{X} R_{\rm WD} / (G M_{\rm WD})$, 
where $L_\mathrm{X}$ is the hard X-ray luminosity, $R_{\rm WD}$ and $M_{\rm WD}$ 
are the radius and mass of the WD, and $G$ is the gravitational constant.
For KPD\,0005+5106's hard X-ray luminosity ($3\times10^{30}$ ergs s$^{-1}$),
mass, and radius, the mass accretion rate would need to be 
2.3$\times$10$^{-12}$ $M_\odot$ yr$^{-1}$, or 1.45$\times$10$^{14}$ g s$^{-1}$.
This mass accretion rate is very small that a Jupiter-like planet donor may survive 
for a few times 10$^8$ yr.

It ought to be noted that KPD\,0005+5106 is a He-rich DO WD, the
accreted material must be He-rich as well; otherwise the WD's atmosphere
will be over-polluted with H.  While a Jupiter-like planet can easily provide
the material to be accreted to power the hard X-ray emission, it is not clear
how such a planet can survive the stellar evolution and what its chemical
composition would be.  It is also not clear how an extremely low-mass star companion
gets so close to a WD whose initial mass may be $\sim$3 $M_\odot$
\citep{Ketal05}.  Binary star evolution needs to be considered for the progenitor 
of this system.

\section{Summary and Conclusion}

Three types of X-ray sources are commonly associated with WDs: soft photospheric
emission, harder X-ray emission from a stellar companion with coronal activity, and 
accretion powered X-ray emission.  It has been puzzling that a small number of 
apparently single WDs are associated with X-ray emission peaking near 1 keV.
The two most conspicuous cases are the central star of the Helix Nebula and 
KPD\,0005+5106.  

With an effective temperature of 200,000 K, KPD\,0005+5106 is a bright soft X-ray
source even detected in the ROSAT All Sky Survey, and a pointed longer
observation reveal an additional harder X-ray component peaking near 1 keV.
We obtained Chandra X-ray Observatory observations of KPD\,0005+5106 
and confirmed the spatial coincidence of the hard and soft X-ray emission
components with the WD.  The very bright soft X-ray emission caused photon pileup, 
rendering the spectral analysis of the hard X-ray emission unreliable.  We have 
thus obtained XMM-Newton X-ray observations of KPD\,0005+5106 and two
other WDs, PG\,1159$-$035, and WD\,0121$-$756, whose ROSAT observations
detected $\sim$10 photons with energies near 1 keV.  The XMM-Newton 
observations show these three WDs have similar X-ray spectral shapes with a bright soft 
component below 0.5 keV and a harder component peaking near 1 keV, and all three
spectra are best fitted by models consisting of a blackbody component for the stellar
photospheric emission, a thermal plasma emission component, and a power-law component.
The hard X-ray luminosities in the 0.6--3.0 keV band are 3.9$\times$10$^{30}$, 3.9$\times$10$^{29}$,
and 3.2$\times$10$^{30}$ ergs s$^{-1}$ for KPD\,0005+5106, PG\,1159$-$035, and 
WD\,0121$-$756, respectively.  

The XMM-Newton EPIC-pn observations of KPD\,0005+5106 detected adequate 
counts for lightcurve analysis.  The lightcurve in the soft energy band (0.3--0.5 keV) is 
essentially constant, but the lightcurve in the hard energy band (0.6--3.0 keV) shows 
variations with an apparent period of $\sim$5 hr.  The hard-band lightcurve of KPD\,0005+5106 
extracted from the Chandra ACIS-S data also shows a similar 
period.  We have thus computed generalized Lomb-Scargle periodograms for the
XMM-Newton and Chandra lightcurves in the hard energy band both
individually and combined.  We find the combined lightcurve, covering 4.6 cycles,
provides the most robust period, 4.7$\pm$0.3 hr, with a false alarm probability
of 0.41\%.

The observed hard X-ray emission cannot arise from a single WD's photosphere.  
Infrared observations of KPD\,0005+5106 place stringent
constraints on possible companions -- must be later and less massive than an M8\,V star. 
Assuming that the 4.7$\pm$0.3 hr
period in the hard X-ray light curve corresponds to a binary orbital period, we considered
a 0.075 $M_\odot$ M9\,V star, a 0.035 $M_\odot$ T brown dwarf, and a 0.001 $M_\odot$
Jupiter-like planet as the companion.   We find the Jupiter-like planet exceeds the Roche
radius and can be a donor providing material to be accreted by the WD to power the hard 
X-ray emission.  The M9\,V star and T brown dwarf, being photoionized and heated by the 
200,000 K KPD\,0005+5106 at a distance of $\sim$1.3 $R_\odot$, may be inflated and 
the outermost material may pass the Roche radius and be accreted by the WD.  
We conclude that the hard X-ray emission from apparently single WDs is powered by
accretion from sub-stellar companions or giant planets, and is modulated by the orbital 
motion with a period of 4.7$\pm$0.3 hr.

\acknowledgments
We thank Dr.\ L.\ Townsley and Dr.\ P.\ Broos for advising us
on the pileup effects in Chandra data, and Dr.\ K.\ Werner for
critically reading this paper. 
This research was supported by the NASA grants
SAO GO8-9026 (Chandra) and JPL 1319342 (Spitzer).
Y.H.C.\ acknowledges grants MOST 108-2112-M-001-045 and 
MOST 109-2112-M-001-040 from the 
Ministry of Science and Technology of Taiwan, Republic of China.
J.A.T.\ and M.A.G.\ are funded by UNAM DGAPA PAPIIT project IA100318.
M.A.G.\ also acknowledges support from grant PGC2018-102184-B-IOO
of the Spanish Ministerio de Ciencia, Educaci\'on y Universidades 
co-funded by FEDER funds.
This paper has used observations made with the Italian Telescopio 
Nazionale Galileo (TNG) operated on the island of La Palma 
by the Fundaci\'on Galileo Galilei of the INAF (Istituto 
Nazionale di Astrofisica) at the Spanish Observatorio del 
Roque de los Muchachos of the Instituto de Astrof\'\i sica 
de Canarias.

\software{CIAO (v4.11; Fruscione et al.\ 2006), SAS (v17.0; Gabriel et al.\ 2004), 
  MOPEX (Makovoz \& Marleau 2005), IRAF (Tody 1986, Tody 1993), XSPEC (v12.10.1; Arnaud 1996)}

\end{document}